\documentclass[%
  aip,
 amsmath,amssymb,
reprint,%
]{revtex4-2}
\usepackage{graphicx}
\usepackage{dcolumn}
\usepackage{bm}
\usepackage{color}
\usepackage{float}

\usepackage[utf8]{inputenc}
\usepackage[T1]{fontenc}
\usepackage{mathptmx}
\usepackage{subfig}
\usepackage[inline]{enumitem}
\usepackage{caption}
\newcommand{\bogus}[1]{{}}

\begin{document}


\title{On the collisional damping of plasma velocity space instabilities}

\author{Yanzeng Zhang and Xian-Zhu Tang}%
\affiliation{Theoretical Division, Los Alamos National Laboratory, Los Alamos, New Mexico 87545, USA}


\begin{abstract}

For plasma velocity space instabilities driven by particle
distributions significantly deviated from a Maxwellian, weak
collisions can damp the instabilities by an amount that is
significantly beyond the collisional rate itself.  This is attributed
to the dual role of collisions that tend to relax the plasma
distribution toward a Maxwellian and to suppress the linearly
perturbed distribution function. The former effect can dominate in
cases where the unstable non-Maxwellian distribution is driven by
collisionless transport on a time scale much shorter than that of
collisions, and the growth rate of the ideal instability has a
sensitive dependence on the distribution function. The whistler
instability driven by electrostatically trapped electrons is used as
an example to elucidate such a strong collisional damping effect of
plasma velocity space instabilities, which is confirmed by
first-principles kinetic simulations.

\end{abstract}

\maketitle

Plasmas of astrophysical, space and laboratory origins are known to
support a wide variety of waves and
instabilities~\cite{stix1992waves,swanson2003plasma,bernstein1958waves,somov2013wave,melrose1986instabilities,chen1987waves}. These waves and instabilities are of great importance for plasma
transport, heating, confinement, and diagnostics.  A large family of
plasma instabilities falls under the category of ideal modes, as they
are excited in the absence of collisions that would introduce
dissipation into an otherwise Hamiltonian system. Plasma collisions
typically reduce the growth rate of these ideal modes, also known by
the term collisional damping in plasma physics. This can be
contrasted with the family of the so-called resistive or dissipative
modes that rely on collisions to destabilize an otherwise stable or
marginal ideal mode. Well-known examples of such include the resistive
tearing modes~\cite{furth1963finite} and a class of dissipative drift
wave instabilities~\cite{hasegawa1983plasma}.  Here we focus on the
collisional damping of ideal modes.

The common expectation is that the collisional damping rate,
$\Gamma_\nu=\gamma_0-\gamma$ with $\gamma_0$ and $\gamma$ the growth
rates without and with collisions, respectively, is approximately the
collisional
rate~\cite{epperlein1992damping,ng1999kinetic,lenard1958plasma,de1999kinetic,brambilla1995effects,aleynikov2015stability},
which for electrons is $\nu=4\sqrt{2\pi}n_e e^4\ln
\Lambda/(3m_e^{1/2}T_e^{3/2})$ with $\ln \Lambda$ the Coulomb
logarithm. This can be understood by considering the Boltzmann
equation
\begin{equation}
   \frac{df}{dt}= \frac{\partial f}{\partial t}+\mathbf{v}\cdot \nabla
   f+\frac{q}{m}\left(\mathbf{E}+\frac{\mathbf{v}\times\mathbf{B}}{c}\right)\cdot\nabla_\mathbf{v}f=C(f),\label{eq_Bolzmann}
\end{equation}
where $C(f)$ is the collision operator.  The stability analysis starts
with identifying an equilibrium distribution $f_0,$ along with the
equilibrium electromagnetic field $(\mathbf{E}_0, \mathbf{B}_0),$ that
satisfies
\begin{align}
\mathbf{v}\cdot\nabla f_0 +
\frac{q}{m}\left(\mathbf{E}_0+\frac{\mathbf{v}\times\mathbf{B}_0}{c}\right)\cdot\nabla_\mathbf{v}f_0
= C(f_0), \label{eq:collisonal-equilibrium-f0}
\end{align}
The perturbed distribution $f_1$ and the perturbed electromagnetic field $(\mathbf{E}_1,\mathbf{B}_1),$
to linear order, follow,
\begin{align}
\frac{\partial f_1}{\partial t}
  & + \mathbf{v}\cdot\nabla f_1
  + \frac{q}{m}\left(\mathbf{E}_0 + \frac{\mathbf{v}\times\mathbf{B}_0}{c}\right)\cdot\nabla_\mathbf{v} f_1 \nonumber
\\
  & + \frac{q}{m}\left(\mathbf{E}_1 + \frac{\mathbf{v}\times\mathbf{B}_1}{c}\right)\cdot\nabla_\mathbf{v} f_0
      = C(f_0+f_1) - C(f_0). \label{eq:df1dt}
\end{align}

The collisional damping of a linearly unstable mode $(f_1, \mathbf{E}_1, \mathbf{B}_1)$ can be understood
by considering a Krook-like approximation of the collision operator~\cite{bhatnagar1954model},
\begin{equation}
    C(f)=-\nu (f-f_{M}),\label{eq_krook-collision}
\end{equation}
which signifies the fact that collisions will relax the particle
distribution function $f$ to a Maxwellian $f_{M}$ over a time period
of $\nu^{-1}$.
This leads to $C(f_0+f_1) - C(f_0) \approx - \nu f_1$ in Eq.~(\ref{eq:df1dt}),
which means that collisions will only affect the linearly
perturbed distribution $f_1$, the effect of which can
be absorbed into the temporal derivative in Eq.~(\ref{eq:df1dt}),
without changing other terms, as
\begin{equation}
   \frac{\partial f_1}{\partial t} +\nu f_1 = -i(\omega+i\nu) f_1,\label{eq-f1}
\end{equation}
where $\omega$ is the mode/wave frequency. The physics implication of
Eq.~(\ref{eq-f1}) is that the collisions will cause damping of the
waves/instabilities by the amount of $\Gamma_\nu=\nu$. It is important
to note that such absorption of $\nu$ into $\omega$ is not applicable
to Maxwell's equations. In the fluid picture, this means that the
substitution $\omega\rightarrow \omega+i\nu$ is only permissible in
the conductivity tensor but not the entire wave dispersion
relation~\cite{aleynikov2015stability}. However, as one can imagine,
even such part substitution would lead to a damping rate
$\Gamma_\nu\sim \nu$, reinforcing the conventional wisdom that the
collisional damping can be important only when $\nu\sim\gamma_0$.

Keeping the $C(f_0)$ term in Eq.~(\ref{eq:collisonal-equilibrium-f0})
implies that the unstable equilibrium distribution $f_0$ is developed
over collisional time scale. This assumption is usually not
  satisfied for ideal modes in which {\em a non-Maxwellian $f_0$ is formed
  and sustained by collisionless transport on a time scale much
  shorter than $1/\nu,$} for which case $C(f_0)$ is absent in
  Eq.~(\ref{eq:collisonal-equilibrium-f0}). Examples of such include
the laboratory sheath/presheath plasmas and tokamak plasmas undergoing
thermal quench that have truncated electron distribution because of
free-streaming
losses~\cite{tang-ppcf-2011,guo2012ambipolar,Zhangfronts}, and coronal
and solar wind plasmas that have strongly anisotropic temperatures,
non-Maxwellian tails, and energetic beam
components~\cite{Marsch-LRiSP-2006,Lazar-etal-FASC-2022}.
For such unstable
collisionless equilibria, the (part) substitution $\omega\rightarrow
\omega+i\nu$ of Eq.~(\ref{eq-f1}) can no longer adequately describe the effect of collisions on the
instability.  This is because weak collisions with $\nu\ll
\gamma_0\ll \omega$ will tend to modify $f_0$ via
\begin{equation}
    \frac{d f_0}{dt}=C(f_0)=-\nu (f_0-f_M) \label{eq:df0dt-by-collision}
\end{equation}
over the same time period in which $f_1$ is collisionally damped as described by
Eq.~(\ref{eq:df1dt}) or Eq.~(\ref{eq-f1}).
Consequently, the collisional modification of $f_0$ by an amount of $\delta f_0$  changes the
linear instability drive in Eq.~(\ref{eq:df1dt}) by the amount of
\[
\frac{q}{m}\left(\mathbf{E}_1+\frac{\mathbf{v}\times\mathbf{B}_1}{c}\right)\cdot\nabla_\mathbf{v}\delta
f_0
\]
on the left-hand side.  In cases where the linear instability has a
sensitive dependence on $\delta f_0,$ which itself grows linearly in
time from Eq.~(\ref{eq:df0dt-by-collision}), the ideal mode can be
collisionally damped mainly via this indirect channel of collisionally modified $f_0.$
For a specific example,
temperature anisotropy can drive a number of instabilities including
the whistler~\cite{kennel1966limit,gary1996whistler},
mirror~\cite{southwood1993mirror,pokhotelov2002linear} and
firehose~\cite{hollweg1970new} instabilities in a magnetized plasma,
and Weibel
instability~\cite{weibel1959spontaneously,kalman1968anisotropic} in an
unmagnetized plasma. For these plasma velocity space instabilities,
the modification of $f_0$ toward $f_M$ by weak collisions can induce
further damping by weakening the drivers (e.g., the temperature
anisotropy) in the velocity space. What is remarkable is that the
resulting {\em collisional damping rate via this indirect route can be much
higher than $\Gamma_\nu\sim\nu$} predicted by the conventional theory
that does not take into account $\delta f_0.$

Having outlined the simple physical picture of such an enhanced
collisional damping mechanism for plasma velocity space instabilities,
we note that its experimental realizability requires two
conditions. The first is that the unstable non-Maxwellian distribution
must be developed on a time scale ($\tau_{trans}$) much faster than
that of the weak collisions ($\tau_c=1/\nu$). This is required to drop
the $C(f_0)$ term in Eq.~(\ref{eq:collisonal-equilibrium-f0}), as
noted earlier. Indeed, for a wide class of problems in the laboratory
and space/astrophysics, the non-Maxwellian distribution is primarily
driven by collisionless transport.  A well-known laboratory example is
the open field line plasma in which free-streaming loss on the time
scale of $L/v_{t}$ sets up a truncated Maxwellian distribution for the
electrons. Here $L$ is the length of the open field line and
$v_{t}=\sqrt{2T_e/m_e}$ is the electron thermal speed
  defined by the electron temperature $T_e$ and electron mass
  $m_e$. If $L$ is much shorter than the electron mean-free-path
$\lambda_{mfp},$ the first condition would be well satisfied. In the
case of solar wind where a range of velocity space instability is
known to exist, the plasma collisionality is tiny and the
non-Maxwellian distribution can be entirely driven by collisionless
transport, for example, as a plasma adiabatic response to magnetic
flux expansion that develops a highly anisotropic distribution
function. We notice there are places where the solar
  wind can be sufficiently isotropic to stay below the
  temperature-anisotropy-driven instability
  threshold~\cite{verscharen2019multi,yoon2019solar}, which could
  result from the collective effect of weak collisions.

The second condition is that the growth rate of the
  ideal instability has a sensitive dependence on the non-Maxwellian
  distribution. This is a necessity for a large impact by the
  collisional modification of $f_0$ that is originally formed by
  collisionless transport on a time scale much shorter than that of
  weak collisions. In other words, since weak collisions can only
  modestly modify $f_0$ on the dynamical time scale of the ideal
  instability, the growth rate of the ideal instability must vary
  significantly with a small change in $f_0$ to observe a large
  effect. Here the dynamical time scale ($\tau_{inst}$) of the
  instability is tied to the linear growth period of the mode, so it
  scales inversely with the linear mode growth rate $\gamma_{0}$
  itself and also logarithmically with how small the initial
  perturbation amplitude ($f_1^i$) is,
  \begin{align}
    \tau_{inst} \approx \frac{1}{\gamma_{0}} \ln \frac{f_1^s}{f_1^i}.
  \end{align}
with $f_1^s$ the perturbation amplitude at the onset of nonlinear
saturation. The first and second conditions combine to imply that in
cases $\tau_{inst}$ is not long compared with $\tau_{trans},$ the $C(f_0)$ term
in Eq.~(\ref{eq:collisonal-equilibrium-f0}), despite its small
amplitude, can already impact the linear mode at its onset due to
latter's sensitivity to the details of $f_0.$

To illustrate the underlying physics, we employ the whistler instability, which is a common velocity space instability in the magnetized plasmas, with the temperature-anisotropy-driven one dated back many decades in space and astrophysics plasmas. Particularly, we focus on the electrostatically trapped electron driven whistler instability~\cite{guo2012ambipolar}, which applies to any magnetized plasma that intercepts a solid wall~\cite{godyak1992measurement,stangeby2000plasma,dorf2009energy}. Notice that we focus on the linear instability and leave the nonlinear saturation physics to future works.  One remarkable property of such whistler instability
is that it has an instability threshold far lower than that of the
whistler instability driven by the temperature
anisotropy~\cite{guo2012ambipolar}.  This was found to be critical for
maintaining ambipolar transport in a steady plasma
(pre)sheath~\cite{guo2012ambipolar} and for cooling the perpendicular
electron temperature in a plasma thermal quench. It is of interest to
note that the thermal quench problem is an ideal motivation for such
an investigation of collisional damping of whistler instability in
that the cooling of the plasma will inevitably bring the plasma from
the initial collisionless regime to the eventual collisional
regime~\cite{LiTQphase}. We will show that a weak collision $\nu\ll \gamma_0$ tends to
smear the trapped-passing boundary in the parallel electron
distribution function, the small modification of which can greatly
reduce or even completely suppress the whistler instability. It must
be emphasized that, although only the trapped electron driven whistler
instability is reported here, other velocity space instabilities like
the whistler and Weibel instabilities driven by the temperature anisotropy
have also been investigated, and the results are in agreement with the
conclusion in this paper.

\begin{figure}
\centering
\includegraphics[width=0.4\textwidth]{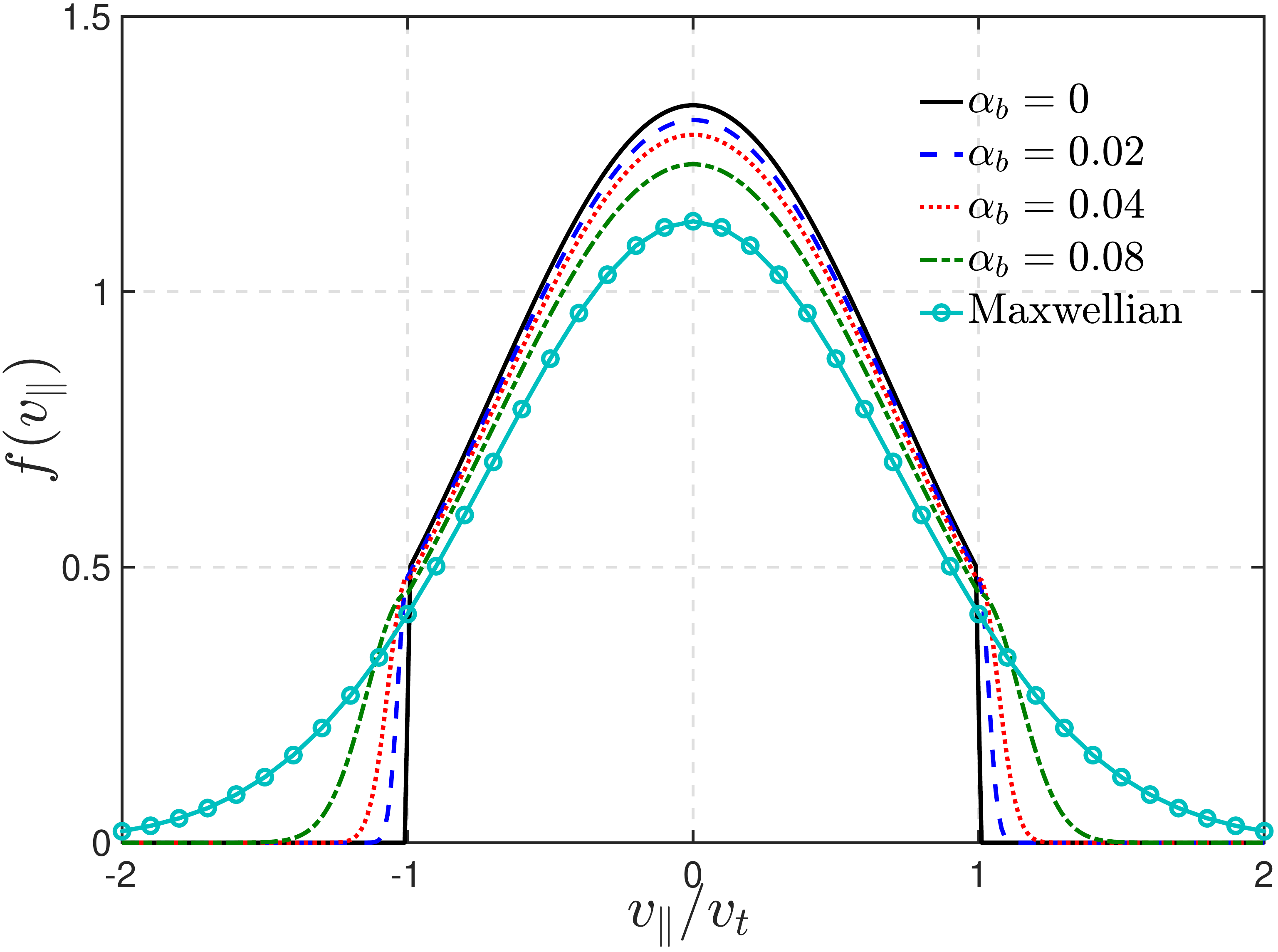}
\caption{Schematic view of $f_e(v_\parallel)$ for $v_c=v_t$. The Maxwellian distribution function with the same density and thermal velocity $v_t$ is shown.}
\label{fig:distrib-smooth-bound}
\end{figure}

Before presenting the first-principles kinetic simulations using
VPIC~\cite{VPIC}, we first employ a simple model to elucidate that
strong damping can be achieved by slightly modifying the distribution
function $f_0$ due to the weak collisions. In the absence of
collisions, the electrostatically trapped electrons can be described
by a cutoff Maxwellian distribution function~\cite{guo2012ambipolar}
\begin{equation}
f_t(v_\parallel,v_\perp)=\frac{2
  n_e}{\textup{Erf}(v_c/v_t)\sqrt{\pi}v_t^3}e^{-(v_\parallel^2+v_\perp^2)/v_t^2}\Theta(1-v_\parallel^2/v_c^2).\label{eq_cutoff_distribution}
\end{equation}
where $\int_{-\infty}^\infty\int_0^\infty f_tv_\perp dv_\perp
dv_\parallel =n_e$ for normalization. Here $v_c\equiv
\sqrt{2e|\phi_{RF}|/m_e}$ is the trapped-passing boundary in
$v_\parallel$, which also affects the height of $f_t(0,0)$ via the
error function $\textup{Erf}(v_c/v_t)$, $\phi_{RF}$ is the reflecting
potential, and $\Theta(x)$ is the Heaviside function. A weak
collision with $\nu\ll \gamma_0$ can affect the whistler instability
through a smeared trapped-passing boundary, as the result of trapped
electrons being scattered into the passing zone through
$C(f_0)$. To see how such smearing of the trapped-passing boundary will
greatly damp the instability, we model $\delta f_0$, due to weak collisions,
with flow-shifted and depleted Maxwellians, which we call electron beams for
convenience,
\begin{equation}
f_b^\pm(v_\parallel,v_\perp)=\frac{4
  n_e}{\sqrt{\pi}v_{t}^2v_{tb}}e^{-v_\perp^2/v_{t}^2}e^{-(v_\parallel\pm
  v_c)^2/v_{tb}^2} \Theta[-(1\pm v_\parallel/v_c)],\label{eq_reflected_distribution-smoother-bound}
\end{equation}
where $\int_{-\infty}^\infty\int_0^\infty f_b^\pm v_\perp dv_\perp
dv_\parallel =n_e$, and $v_{tb}$ denotes the width of $f_b^\pm$ in
$v_\parallel$ and hence the degree of smoothness of the total distribution. Notice that such
choice of $f_b^\pm$ enables us to take advantage of the incomplete
plasma dispersion
function induced
by a depleted Maxwellian~\cite{franklin1971proceedings,baalrud2013incomplete}. As a result, the trapped electron
distribution function after the smearing of trapped-passing boundary by
the collisions can be modeled as
\begin{equation}
f_e=(1-\alpha_b)f_t+\alpha_b (f_b^++f_b^-)/2,\label{eq-distri-smoother-bound}
\end{equation}
where $\alpha_b$ is the fraction of beam electron density. For a
smooth transition of $f_e$ at $\pm v_c$, we will take $(1-\alpha_b)
f_t=\alpha_b f_b^\pm/2$ at $v_\parallel=\pm v_c$ by using the proper
$v_{tb}=\alpha_b v_t \textup{Erf}(v_c/v_t)
\exp(v_c^2/v_t^2)/(1-\alpha_b)$. This illustrates that the larger
fraction of electron beams $\alpha_b$ will cause a smoother
trapped-passing boundary (larger $v_{tb}$). As an example, $f_e$ for
different $\alpha_b$ (and thus $v_{tb}$) at $v_c=v_t$ are shown in
Fig.~\ref{fig:distrib-smooth-bound}.
 
The distribution function $f_e$ in
Eq.~(\ref{eq-distri-smoother-bound}) should be placed in the
dispersion relation of a whistler wave propagating along a uniform
background magnetic field~\cite{Krallpriciples} with normal mode \textit{ansatz}
$\exp(ikx_\parallel-i\omega t)$,
\begin{align}
1-\frac{k^2c^2}{\omega^2}+&\frac{\omega_{pe}^2}{n_e\omega}\int_{-\infty}^\infty\int_0^\infty\Big[\left(1-\frac{kv_\parallel}{\omega}\right)\frac{\partial
    f_e}{\partial v_\perp^2}
  \nonumber\\
  &+\frac{kv_\parallel}{\omega}\frac{\partial
    f_e}{\partial
    v_\parallel^2}\Big]\frac{v_\perp^3}{\omega-kv_\parallel
  -\omega_{ce}}dv_\perp dv_\parallel=0,\label{eq-dispersion-def}
\end{align}
where we have ignored the effect of ions assuming $\omega_{ci}\ll
\omega<\omega_{ce}$, $\omega_{pe}$ is the plasma frequency,
$\omega_{ce,i}$ is the electron (ion) gyro-frequency, and $c$ is the
speed of light in a vacuum. As a result, the dispersion relation is
given by
\begin{equation}
D(\omega,k)=1-\frac{k^2c^2}{\omega^2}+(1-\alpha_b)D_t+\alpha_b
D_b=0,\label{eq-dispersion-trap-smooth-bound}
\end{equation}
where $D_t$ and $D_b$ are from the trapped and beam electrons,
respectively
\begin{align}
D_t&=\frac{\omega_{pe}^2}{\textup{Erf}(\hat{v}_c)\sqrt{\pi}\omega^2}\left[\frac{\omega}{kv_t}\int_{-\hat{v}_c}^{\hat{v}_c}\frac{e^{-\hat{v}_\parallel^2}}{\hat{v}_\parallel-\xi}d\hat{v}_\parallel
  +\frac{\hat{v}_ce^{-\hat{v}_c^2}}{\hat{v}_c^2-\xi^2}\right],\label{eq-dispersion-trap}
  \end{align}
\begin{align}
D_b&=\frac{\omega_{pe}^2}{\omega^2}\Big[ -1+\frac{1}{\hat{v}_{tb}^2}-\frac{1}{\sqrt{\pi}} \frac{\hat{v}_c/\hat{v}_{tb}}{\hat{v}_c^2-\xi^2}+\nonumber\\
&\sum_\pm \pm\frac{\hat{\omega}_{ce}+(\xi\pm \hat{v}_c)/\hat{v}_{tb}^2}{\sqrt{\pi} \hat{v}_{tb}}\int_{\mp\infty}^0 \frac{e^{-\eta^2}}{\eta-(\xi\pm \hat{v}_c)/\hat{v}_{tb}}d\eta\Big],\label{eq-dispersion-smooter-bound-beam}
\end{align}
$\hat{v}_{\parallel, c,tb}=v_{\parallel, c,tb}/v_t$,
$\hat{\omega}_{ce}=\omega_{ce}/kv_t$, $\xi=(\omega-\omega_{ce})/kv_t$
and $\eta=(v_\parallel\pm v_c)/v_{tb}$. The integrals in
Eqs.~(\ref{eq-dispersion-trap},
\ref{eq-dispersion-smooter-bound-beam}) can be evaluated using the
incomplete plasma dispersion
function~\cite{baalrud2013incomplete,franklin1971proceedings},
$Z(w,u)$, which is similar to the plasma dispersion
function~\cite{Frieddispersion}, $Z(w)$, but has a cutoff at the lower
limit of the integral, $u$.

In the absence of $f_b^\pm$, the most unstable mode (or resonant
condition) satisfies $\omega_r-\omega_{ce}\approx \pm kv_c$ (i.e.,
$\xi_r \approx \pm \hat{v}_c$) for whistler waves with
$\omega_r<\omega_{ce}$ as seen from $D_t$ since there is no
counterpart with $|v_\parallel| > v_c$ in $f_t$. This means that the
resonant electrons have parallel velocity $v_\parallel \approx \pm
v_c$ for whistler modes with $\mp k>0$. However, to avoid damping
in the integral induced by the singular pole along Landau-like
contour, this mode will have $|\xi|_r$ slightly greater than
$\hat{v}_c$.

When there is a smoother boundary with $f_e(|v_\parallel|>v_c)> 0$,
the resonant condition can be modified, leading to a reduction of the
growth rate. Many physical insights into the impact of a smoother trapped-passing
boundary on whistler instability can be obtained from the analytical
solution of Eq.~(\ref{eq-dispersion-trap-smooth-bound}) in two
limiting cases. The first one is a small cutoff speed, $v_c\ll
v_t$. Notice that $\omega_r-\omega_{ce}\approx kv_c \ll k v_t$
provides $\omega_r\approx \omega_{ce}$. Under such a condition, both
$\xi+ \hat{v}_c$ and $\xi-\hat{ v}_c$ can contribute equally to $D$
when $\xi\approx i\gamma/(kv_t)>\hat{v}_c$. Moreover, since
$v_{tb}\sim \alpha_b v_c/(1-\alpha_b)\ll v_c$ for small $\alpha_b$,
the sum of the integrals in $D_b $ are approximated to the plasma
dispersion function with a large argument. As a result, if
$|\omega/( kv_t)| \ll |1/\xi|$, one finds in the limit of $k^2c^2\gg \omega^2$ 
that
\begin{equation}
D=-\frac{k^2c^2}{\omega^2}+(1-\alpha_b)\frac{\omega_{pe}^2}{2\omega^2}\frac{k^2v_t^2}{\gamma^2}+\alpha_b \frac{\omega_{pe}^2}{2\omega^2}\frac{k^2v_t^2}{\gamma^2},\label{eq_dispersion_small_vc_smooth_bound}
\end{equation}
where the second (third) term is from the trapped electrons (electron
beams), and the approximation of
$\textup{Erf}(x)\rightarrow 1.125 x$ is invoked for $x\ll
1$. Eq.~(\ref{eq_dispersion_small_vc_smooth_bound}) illustrates that
the electron beams do not affect the whistler instability for $v_c\ll
v_t$, yielding a solution $\omega\approx
\omega_{ce}+i\omega_{pe}v_{t}/\sqrt{2}c$. This is not surprising
considering that a delta-function-like profile of $f_b^\pm$ at
$v_{tb}\ll v_c\ll v_t$ does not introduce appreciable smoothing.

For a general cutoff velocity, $v_c\sim v_t$, we can consider a small
fraction of electron beams $\alpha_b\ll 1$ (weak smoothing) so that
$\hat{v}_{tb}\ll |\xi\pm \hat{v}_c|$. Notice that for the most
unstable mode with $k>0$ ($k<0$), there is only one resonant condition
$\xi=-\hat{v}_c$ ($\xi=\hat{v}_c$). As a result, the incomplete plasma
dispersion functions in $D_b$ can be approximated by an asymptotic
expansion of the large argument to find,
\begin{equation}
D_b \approx
\frac{\omega_{pe}^2}{\omega^2}\Big[-1+\frac{\hat{\omega}_{ce}\xi}{\hat{v}_c^2-\xi^2}-\frac{1}{2}\frac{\xi^2+\hat{v}_c^2}{(\xi^2-\hat{v}_c^2)^2}\Big].\label{eq-Db_small-vtpara}
\end{equation}
This approximation is also applicable to the small $v_c$ limit, where
the third term in the bracket dominates for small $\xi,$ providing the
third term in Eq.~(\ref{eq_dispersion_small_vc_smooth_bound}). While $D_t$ can be approximated as
\begin{align}
D_t&\approx\frac{\omega_{pe}^2}{\omega^2}
\Big\{\frac{\omega_r}{kv_c}+i\Big[ \frac{\omega_r}{(kv_c)^2} \gamma -
  \frac{kv_te^{-\hat{v}_c^2}}{2\sqrt{\pi} \gamma}
  \Big]\Big\}.\label{eq-dispersion-trap-large-vc-smooth-bound}
\end{align}
In the limit of $k^2c^2\gg \omega^2$, the growth rate is given by
$(1-\alpha_b)Im(D_t)+\alpha_b Im(D_b)=0$, where the factor
$\omega_{pe}^2/\omega^2$ has been excluded from $D$. This yields
\begin{equation}
(1-\alpha_b)\Big[\frac{\omega_r}{(kv_c)^2} \gamma -
    \frac{kv_te^{-\hat{v}_c^2}}{2\sqrt{\pi} \gamma}
    \Big]+\alpha_b\frac{\omega_{ce}}{2\gamma}=0,\label{eq-imag-part-smooth-bound}
\end{equation}
where the resonant condition is assumed to be exactly $|\xi_r|=\hat{v}_c$
and thus the third term in Eq.~(\ref{eq-Db_small-vtpara}) can be
ignored for the imaginary part. Eq.~(\ref{eq-imag-part-smooth-bound})
indicates that the electron beams will reduce the whistler instability
and even completely suppress it when $\alpha_b\gtrsim
\alpha_b^{th}\equiv kv_t exp(-\hat{v}_c^2)/(\sqrt{\pi}\omega_{ce})$.

\begin{figure}
\centering
\includegraphics[width=0.35\textwidth]{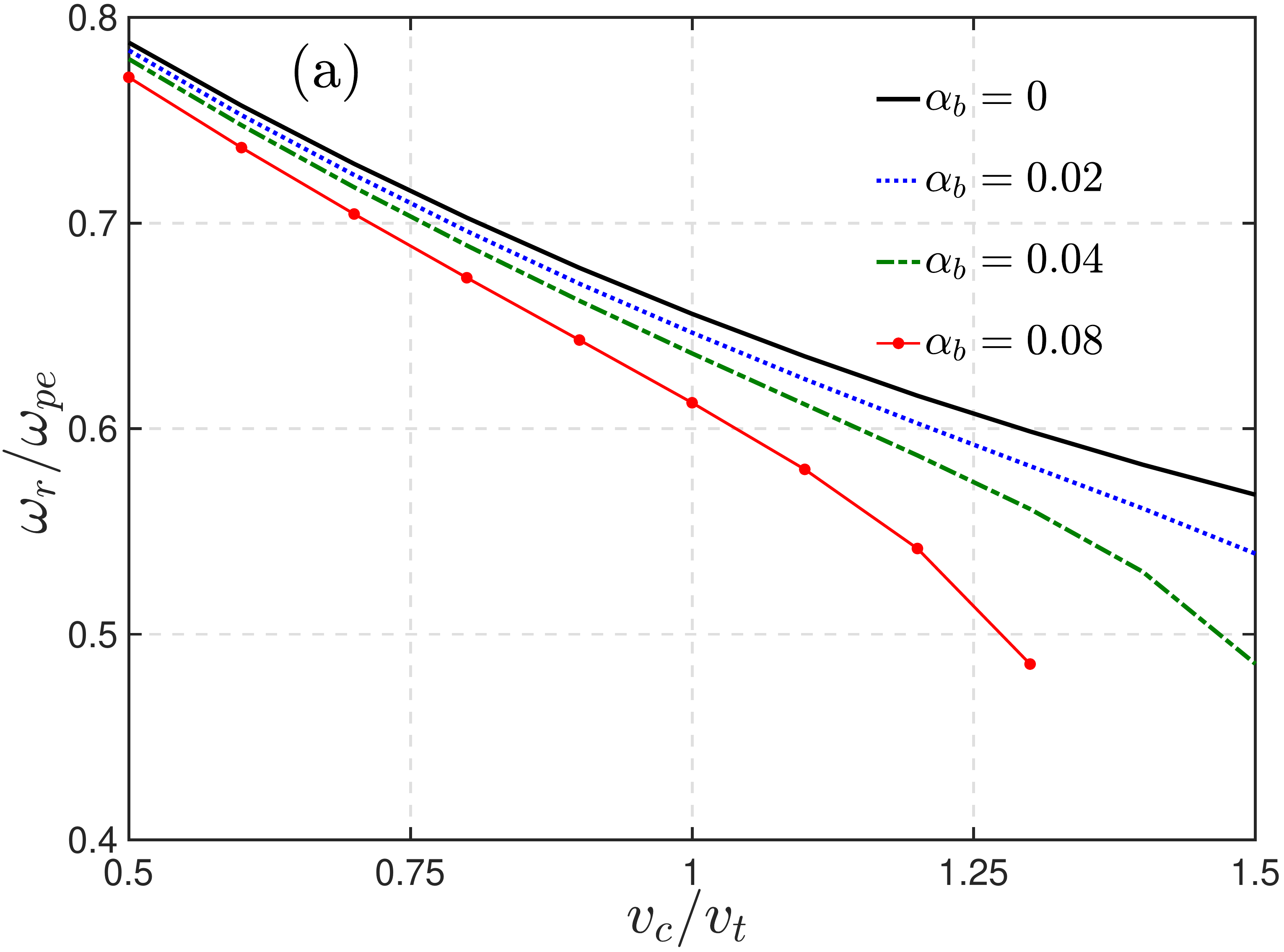}
\includegraphics[width=0.35\textwidth]{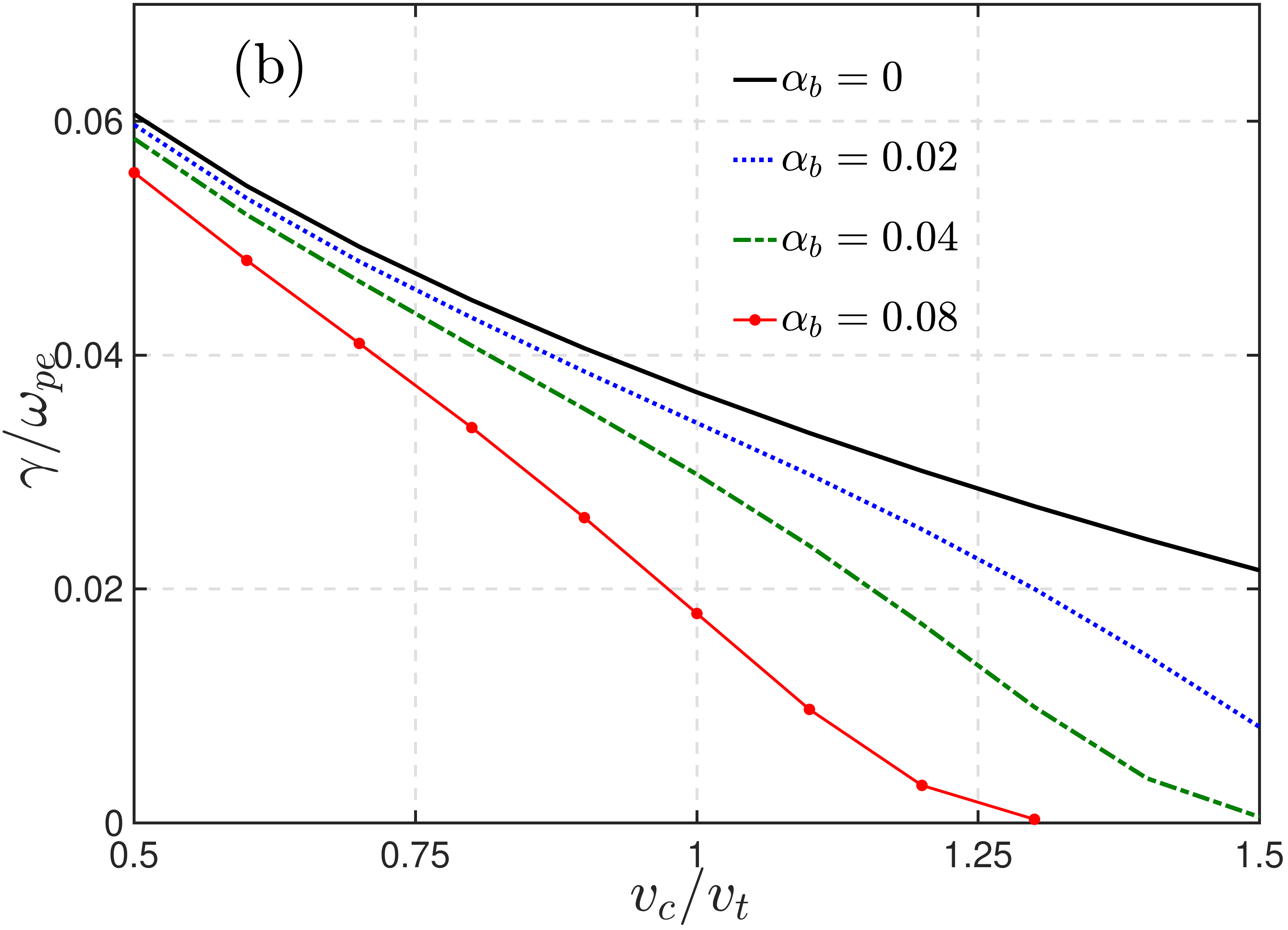}
\caption{Real frequency (a) and growth rate (b) of the most unstable mode for different $\alpha_b$ corresponding to figure~\ref{fig:distrib-smooth-bound}. The plasma parameters are
chosen from a thermal quench problem of a fusion-grade plasma with
density $n_e=10^{19}m^{-3}$, temperature $T_e=T_i= 10keV$, and an external
magnetic field, $B_0$ so that $\beta_e\equiv 8\pi
n_eT_e/B_0^2=4\%$.}
\label{fig:growth-simulation}
\end{figure}

For a thermal quench problem, the numerical solutions of Eq.~(\ref{eq-dispersion-trap-smooth-bound}) are
plotted in Fig.~\ref{fig:growth-simulation} for the real frequency and
growth rate of the most unstable mode. It shows that for $v_c\sim v_t$, the growth
rate is greatly reduced by a smoother trapped-passing
boundary. Such a reduction is more
significant for larger $v_c$ and $\alpha_b$ as suggested by
Eq.~(\ref{eq-imag-part-smooth-bound}). For sufficiently large
$\alpha_b$, which depends on $v_c$, the instability can be
completely suppressed. In contrast, for small $v_c\ll v_t$, the damping
effect is small (and negligible for $v_c\lesssim 0.1 v_t$ from
numerical solutions that are not shown), in good agreement with
Eq.~(\ref{eq_dispersion_small_vc_smooth_bound}).

Fig.~\ref{fig:growth-simulation} demonstrates that the smearing
of the trapped-passing boundary will also reduce the real frequency and
hence the wavenumber of the whistler mode. This is because the resonant
condition $\omega-kv_\parallel=\omega_{ce}$ will cover larger
$v_\parallel>v_c$ and hence smaller $\omega$ and $k$ with smoother
trapped-passing boundary. Such property indicates that the damping of
velocity space instabilities due to $C(f_0)$ is realized through
changing the resonant condition due to the modification of the
distribution function $f_0$.

Although such model analysis confirms that moderate smearing of the
trapped-passing boundary, which is physically due to weak collisions
where $\alpha_b$ increases with the collisional rate $\nu$, can cause
appreciable damping ($\Gamma_\nu=\gamma_0-\gamma\sim \gamma_0$) of the
whistler instability, the quantification of $\Gamma_\nu$ with $\nu$
can only be obtained by deploying the first-principles kinetic
simulations. Here we employ 1D3V PIC simulations using the VPIC code~\cite{VPIC}, which is relativistic, to investigate the whistler instability driven by the trapped electrons. A uniform plasma with parameters corresponding to Fig.~\ref{fig:growth-simulation} is initiated in a periodic box with a length of $L_x=1400\lambda_D$ with $\lambda_D$ being the Debye length. The
ion distribution function is a Maxwellian but electrons are drawn from $f_t$ in Eq.~(\ref{eq_cutoff_distribution}). Reduced ion mass $m_i=100m_e$ is used. The resolution of the simulation is $\Delta x
  =0.1\lambda_D$ with 5000 macro-particles per cell. Takizuka and Abe's method~\cite{T.A} is employed as the collisional
  model in VPIC, where we vary the collisional rate by utilizing an artificial Coulomb logarithm $\ln \Lambda$.  
  Notice that the most unstable mode will arise from the incoherent thermal
noise in VPIC and become dominant over time.
  The key idea for
such simulations is that even weak collisions $\nu\ll \gamma_0\sim
10^{-2}\omega_{pe}$ can cause smearing of the trapped-passing boundary in
the time period of $\gamma_0^{-1}$ so that the linear growth rate will
decrease with time compared to that for the collisionless case.

\begin{figure}
\centering
\includegraphics[width=0.35\textwidth]{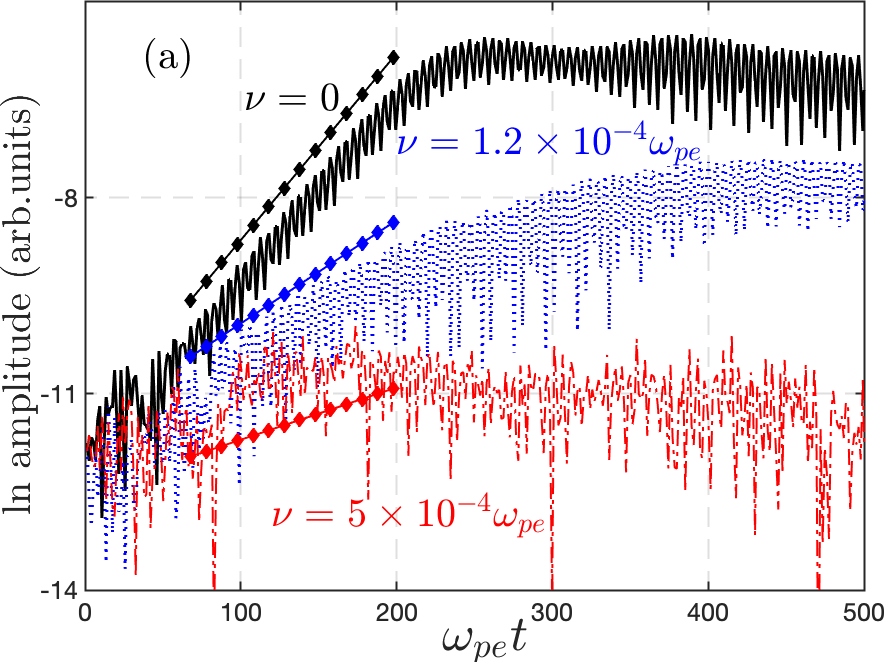}
\includegraphics[width=0.35\textwidth]{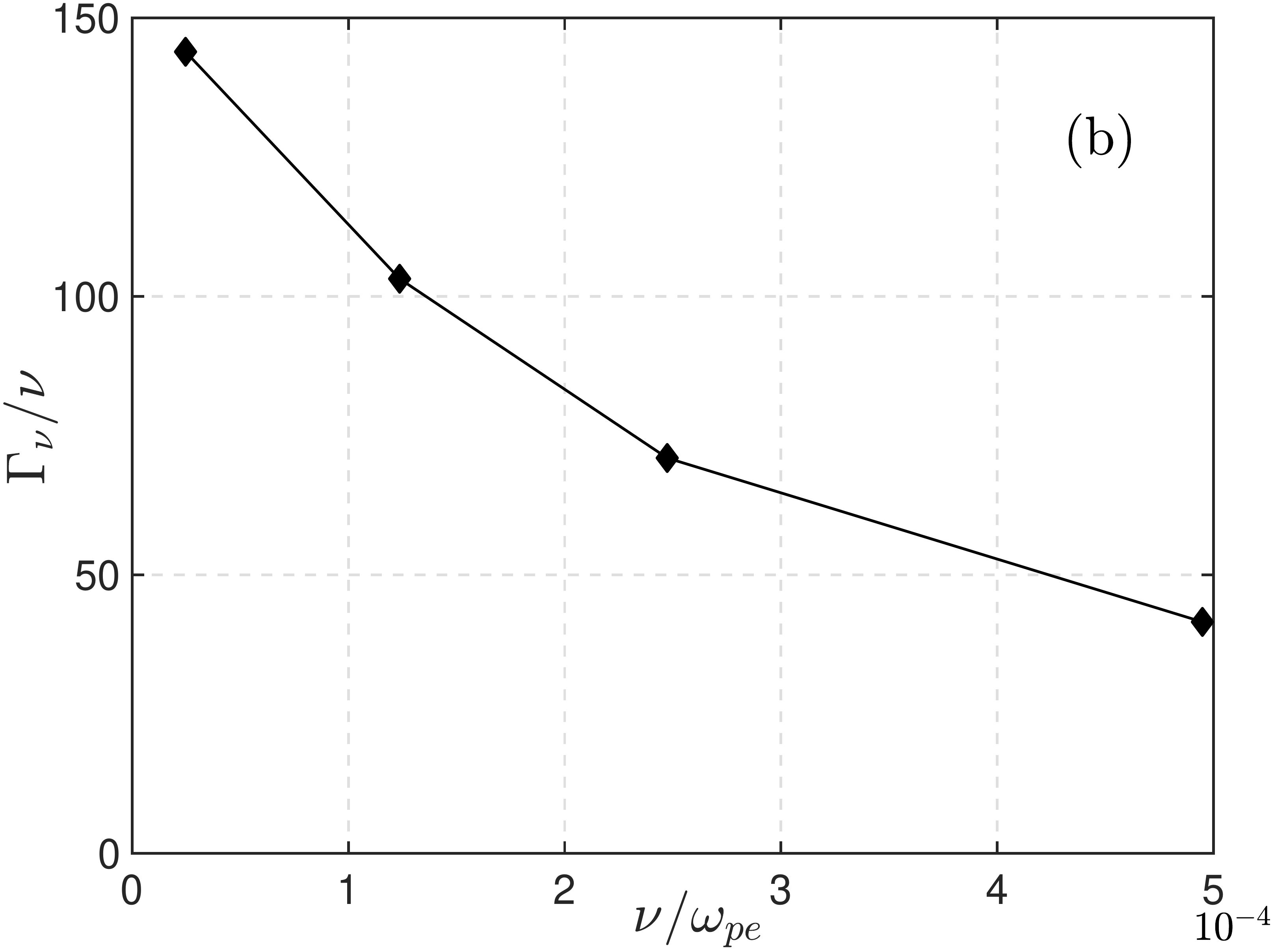}
\caption{(a) Time evolution of the amplitude of perturbed magnetic field \textcolor{red}{(in natural-log scale)} for the most unstable mode (the solid lines labeled by diamonds illustrate the fitted slope). and (b) the collisional damping rate versus the collisional rate for $v_c=v_t$. }
\label{fig:whistler_coll}
\end{figure}

\begin{figure}
\centering
\includegraphics[width=0.35\textwidth]{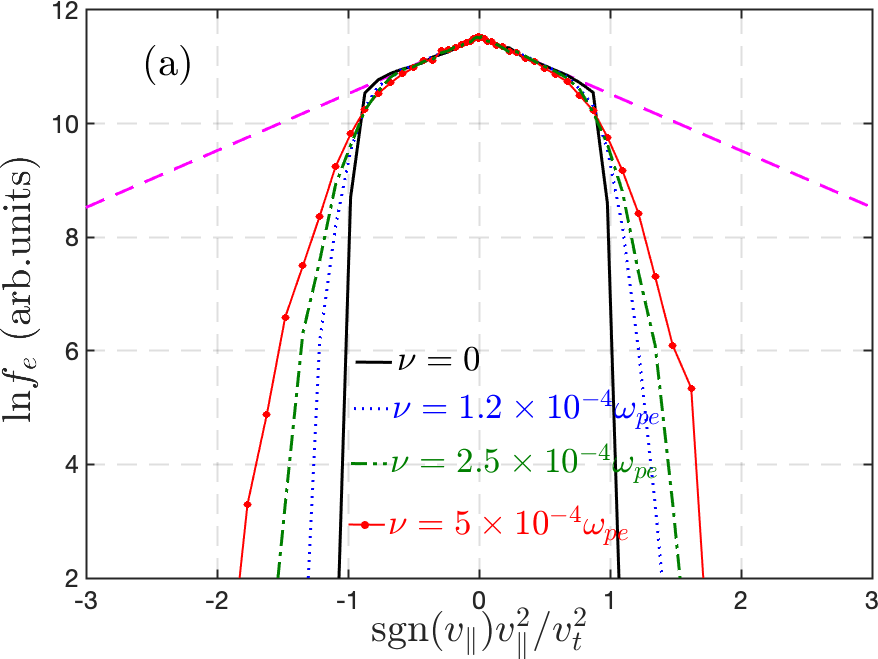}
\includegraphics[width=0.35\textwidth]{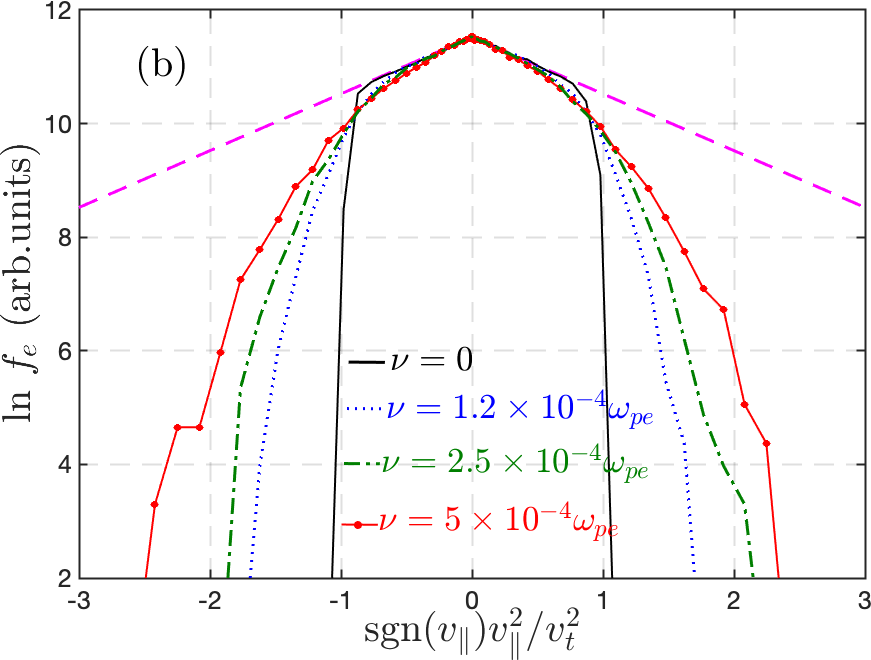} 
\caption{Parallel electron distribution functions at $\omega_{pe}
  t\approx 41$ and (b) $\omega_{pe} t\approx 150$ for different $\nu$
  corresponding Fig.~\ref{fig:whistler_coll}. The dashed line
  (magenta) represents the Maxwellian distribution function.}
\label{fig:whistler-distribution}
\end{figure}

In Fig.~\ref{fig:whistler_coll}, we show the time evolution of the
amplitude of the perturbed transverse magnetic field for the most
unstable mode with an initial cutoff velocity of $v_c=v_t$. In the
absence of collisions, the electron distribution function in the
linear regime remains a cutoff Maxwellian (e.g., see
Fig.~\ref{fig:whistler-distribution}). As a result, the growth rate
remains nearly constant (e.g., see Fig.~\ref{fig:whistler_coll}), which is fitted as
$\gamma_0=0.029\omega_{pe}$. We notice that such a growth rate is smaller than
  the analytical result in Fig.~\ref{fig:growth-simulation} with $\alpha =0$, where
  $\gamma_0=0.036\omega_{pe}$. This is because the distribution in
  VPIC cannot sustain a cutoff Maxwellian with a discontinuity, so the
  trapped-passing boundary will be slightly smoothed as shown in
  Fig.~4 starting from the first-step advancement of the
  simulations. As a result, the growth rate should be smaller than
  that of an exact cutoff Maxwellian, reinforcing the observation that the linear growth rate
  has a sensitive dependence on the fine details of the distribution function.
  In fact, if we integrate the numerical distribution from VPIC to the
  dispersion relation in Eq.~(\ref{eq-dispersion-def}), we obtain
  $\gamma_0=0.03\omega_{pe}$, agreeing well with the fitted growth
  rate from Fig.~\ref{fig:whistler_coll}. 

Fig.~\ref{fig:whistler_coll} shows that even weak collisions $\nu\ll \gamma_0$ will
continuously smear the trapped-passing boundary (e.g., see
Fig.~\ref{fig:whistler-distribution}), which, according to our model
analysis, will cause increasing damping of the linear instability with
time. For such cases, there are two regimes concerning the collisional
rate: (1) if the collision is so weak that the modification of $f_t$
before nonlinear saturation is moderate, the whistler modes keep
growing with decreasing growth rate; and (2) if the collisional rate
is relatively large, the smearing of the trapped-passing boundary will
reach the point that all unstable modes are suppressed. These two
regimes have been illustrated in Fig.~\ref{fig:whistler_coll}(a),
where the transition of them occurs at $\nu\sim 10^{-4}\omega_{pe}\ll
\gamma_0$.
 
The growth rates of whistler instability with collisions are fitted by
avoiding the early time $\omega_{pe}t>100$ to illustrate the strong
collisional damping effect as shown in
Fig.~\ref{fig:whistler_coll}(a), from which we have plotted
$\Gamma_\nu/\nu$ versus $\nu$ in Fig.~\ref{fig:whistler_coll}(b). A
remarkable finding is that $\Gamma_\nu$ is two orders of magnitude
larger than $\nu$, which is much stronger than the conventional theory
of collisional damping rate with $\Gamma_\nu/\nu\sim
1$. It is interesting to note that $\gamma\sim
   10^{-2}\omega_{pe}$ for the whistler instability and thus
   appreciable collisional damping of the whistler instability requires
   $\nu\sim 10^{-4}\omega_{pe}$. In such a regime, the condition
   that $f_0$ is determined by the hot tail electron loss mechanism
   instead of collisions, i.e., $\tau_{trans}\sim L/v_c\ll \tau_c\sim 1/\nu$,
   requiring $L/v_c\sim 10^{3}\omega_{pe}^{-1}$ or $L\sim
   10^{3}\lambda_D$, which can be satisfied in magnetized plasmas that
   intercept a solid wall.

In conclusion, we have shown that a drastically enhanced collisional damping is realized for the ideal mode driven by an
equilibrium $f_0$ reached in the collisionless limit that is significantly deviated from a Maxwellian. Such stronger collisional damping is due to the modification of $f_0$ via collision operator $C(f_0)$ as opposed to the damping of $f_1$ via $C(f_1)$. An example
of the trapped electron driven whistler instability is
used to elucidate such a mechanism, where weak collisions can cause a
significant damping of the instability by smearing the trapped-passing
boundary. The first-principles simulations show that
$\Gamma_\nu/\nu\sim 10^2$, which is much beyond the conventional
theory when $\delta f_0$ is not taken into account.

We thank the U.S. Department of Energy Office of Fusion Energy
Sciences and Office of Advanced Scientific Computing Research
for support under the Tokamak Disruption Simulation (TDS) Scientific
Discovery through Advanced Computing (SciDAC) project, and the Base
Theory Program, both at Los Alamos National Laboratory (LANL) under
contract No. 89233218CNA000001. Y.Z. is supported under a
Director’s Postdoctoral Fellowship at LANL. This research used
resources of the National Energy Research Scientific Computing Center
(NERSC), a U.S. Department of Energy Office of Science User Facility
operated under Contract No. DE-AC02-05CH11231 and the Los Alamos
National Laboratory Institutional Computing Program, which is
supported by the U.S. Department of Energy National Nuclear Security
Administration under Contract No. 89233218CNA000001.

\bibliography{reference}

\end{document}